\documentclass[twocolumn,prl,aps,showpacs]{revtex4}
\usepackage{epsfig}

\begin{document}

\title{First Principles Calculation of Anomalous Hall Conductivity in Ferromagnetic bcc Fe}
\author{Yugui Yao$^{1,2,3}$, L. Kleinman$^{1}$, A. H. MacDonald$^{1}$, Jairo Sinova$
^{4,1}$, T. Jungwirth$^{5,1}$, Ding-sheng Wang$^{3}$, Enge Wang$^{2,3}$, Qian Niu$^{1}$}

\affiliation{$^1$Department of Physics, University of Texas, Austin, Texas 78712} 
\affiliation{$^2$International Center for Quantum Structure, Chinese Academy of Sciences, Beijing 100080, China}
\affiliation{$^3$Institute of Physics, Chinese Academy of Sciences, Beijing 100080, China}
\affiliation{$^4$Department of Physics, Texas A\&M University, College Station, TX 77843-4242}
\affiliation{$^4$ Institute of Physics  ASCR, Cukrovarnick\'a 10, 162 53
Praha 6, Czech Republic }
\date{\today}


\begin{abstract}
We perform a first principles calculation of the anomalous Hall effect in ferromagnetic bcc Fe.
Our theory identifies an intrinsic contribution to the anomalous Hall conductivity and 
relates it to the k-space Berry phase of occupied Bloch states.  The theory is able to 
account for both dc and magneto-optical Hall conductivities with no adjustable parameters.  
\end{abstract}
\pacs{75.47.-m, 71.15.-m, 72.15.Eb, 78.20.Ls}
\maketitle

In many ferromagnets the Hall resistivity, $\rho _H$, exhibits 
an anomalous contribution proportional to the magnetization of the material, in 
addition to the ordinary contribution proportional to the applied magnetic field,
$\rho _{_H}=R_0B+R_s4\pi M$ \cite{smith1929,chien1980,hall1880}. 
The anomalous Hall effect (AHE) has played an important role in the 
investigation and characterization of itinerant electron ferromagnets  
because $R_s$ is usually at least one order of magnitude larger than
the ordinary constant $R_0$. 
Although the effect has been recognized for more than a century \cite{hall1880}
it is still somewhat poorly understood, a circumstance reflected by the   
controversial and sometimes confusing literature on the subject. 
Previous theoretical work has failed to explain the magnitude of the 
observed effect even in well understood materials like Fe \cite{leribaux1966}.   

Karplus and Luttinger \cite{luttinger1954} pioneered the theoretical investigation of this effect,
by showing how spin-orbit coupling in Bloch bands can give rise to an anomalous Hall conductivity (AHC)
in ferromagnetic crystals.  Their conclusion was questioned by Smit \cite{smit1955}, 
who argued that $R_s$ must vanish in a periodic lattice. Smit proposed an alternative mechanism, skew scattering, 
in which spin-orbit coupling causes spin polarized electrons to be scattered preferentially to 
one side by impurities.  The skew scattering mechanism predicts an anomalous Hall 
resistivity linearly proportional to the longitudinal resistivity; this is 
in accord with experiment in some cases, but an approximately quadratic proportionality
is more common.  Later, Berger \cite{berger1970} proposed yet 
another mechanism, the side jump, in which the trajectories of scattered electrons shift to 
one side at impurity sites because of spin-orbit coupling.  The side jump mechanism does predict
a quadratic dependence of the AHC on the longitudinal resistivity. 
However, because of inevitable difficulties in modeling impurity scattering
in real materials, it has not been possible to compare quantitatively with experiment.
It appears to us that the AHE has generally been regarded as an extrinsic
effect due solely to impurity scattering, even though this notion has never been
critically tested, and that the intrinsic contribution initially 
proposed by Karplus and Luttinger has been discounted.

Several years ago, the scattering free contribution of Karplus and Luttinger was 
rederived in a semiclassical framework of wavepacket motion in Bloch bands by taking 
into account Berry phase effects \cite{niu1,niu2}.  
According to this work, the AHC in the scattering free limit is 
a sum of Berry curvatures (see eqs.(\ref{2}) and (\ref{7}) below) of the occupied Bloch states
\cite{berry}.  Recently, Jungwirth \textit{et al.} \cite{jungwirth2002,jungwirth2003} 
applied this picture of the  AHE to (III,Mn)V ferromagnetic semiconductors 
and found very good agreement with experiment.  (III,Mn)V  
ferromagnets are unusual, however, because they are strongly disordered and have
extremely strong spin-orbit interactions.  
In this Letter, we report on an evaluation of the intrinsic AHC in a classic
transition metal ferromagnet, bcc Fe.  Our calculation is
based on spin-density functional theory and the LAPW method. 
The close agreement between theory and experiment that we find leads us to conclude
that the AHC in transition metal ferromagnets is intrinsic in origin, except
possibly at low-temperature in highly conductive samples. 
   
We begin our discussion by briefly reviewing the semi-classical transport theory. 
By including the Berry phase correction to the group velocity \cite{niu1, niu2},
one can derive the following equations of motion: 
\begin{eqnarray}
\stackrel{\cdot }{\mathbf{x}}_c &=& \frac 1\hbar \frac{\partial \varepsilon _n({\bf k})}{%
\partial \mathbf{k}}-\stackrel{\cdot }{\mathbf{k}}\times \mathbf{\Omega }_n(\mathbf{k}),   \label{1} \\
\hbar \stackrel{\mathbf{.}}{\mathbf{k}} &=& -e\mathbf{E}-e\stackrel{\cdot }{%
\mathbf{x}_c}\times \mathbf{B},  \nonumber
\end{eqnarray}
where $\mathbf{\Omega }_n$ is the Berry curvature of the Bloch state defined by 
\begin{equation}
\mathbf{\Omega }_n( \mathbf{k}) =-\mathop{\rm Im}\left\langle
\nabla _{\mathbf{k}}u_{n\mathbf{k}}\left| \times \right| \nabla _{\mathbf{k}%
}u_{n\mathbf{k}}\right\rangle,   \label{2}
\end{equation}
with $u_{n\mathbf{k}}$ being the periodic part of the Bloch wave in the $n$th band. We will be interested in the case
of $\mathbf{B}=0$, for which $\varepsilon _n({\bf k})$ is just the band energy.
The distribution function satisfies the Boltzmann equation with the usual drift and scattering terms, and can 
be written as $f_n(\mathbf{k}) +\delta f_n( \mathbf{k}) $, where $f_n$ is the equilibrium Fermi-Dirac distribution function and $\delta f_n$ is a shift proportional to the electric field and relaxation time.  The electric current is given by the average of the velocity over the distribution function, yielding to first order in the electric field \cite{luttinger1958} 
\begin{equation}
-\frac{e^2}\hbar \mathbf{E}\times \int d^3{k \sum_{n}f_n\mathbf{\Omega}_n}(\mathbf{k})
-\frac e\hbar \int d^3{k}\, \sum _n\delta f_n(\mathbf{k}) \frac{\partial \varepsilon _n 
}{\partial \mathbf{k}}.  \label{4}
\end{equation}
The same expressions can be derived from the Kubo linear-response-theory formula for the conductivity matrix.
The first term is the anomalous Hall current originally derived by Karplus and Luttinger, but never 
previously evaluated.  In the second term, apart from the longitudinal current, there can also be a Hall current in the
presence of skew scattering because the distribution function can acquire a shift in the
transverse direction.  This {\em skew scattering} contribution
to the Hall conductivity should be much smaller than, but crudely proportional to, the longitudinal conductivity and can be 
identified, when it is dominant, by the traditional test, {\it i.e.} $\rho_{xy} \propto \rho_{xx}$.  It will have 
a larger relative importance when $\sigma_{xx}$ is large, {\it i.e.} in pure-crystals at low temperatures.

We now discuss our scheme for calculating the Berry curvature and the AHC. 
For a cubic material with magnetization aligned along %
\mbox{$[$}001\mbox{$]$}, only the $z$-component $\Omega 
^{z}( \mathbf{k}) \neq 0$. 
In our calculation, we find it convenient to use a different but equivalent expression for the Berry curvature 
that arises naturally from the Kubo formula derivation \cite{thouless1982},
\begin{equation}
\Omega ^z _n( \mathbf{k}) =-\sum_{n^{\prime } \neq n}\frac{2%
\mathop{\rm Im}\left\langle \psi _{n\mathbf{k}}\left| v%
_x\right| \psi _{n^{\prime }\mathbf{k}}\right\rangle \left\langle \psi
_{n^{\prime }\mathbf{k}}\left| v_y\right| \psi _{n\mathbf{k}%
}\right\rangle }{( \omega _{n^{\prime }}-\omega _n) ^2},  \label{5}
\end{equation}
where $E_n=\hbar \omega _n,$ and $v$'s are velocity operators.  
In the relativistic formulation, the $\psi _{n\mathbf{k}}$ are
four-component Bloch wave functions, and ${\bf v}= c ( 
\begin{array}{ll}
0 & \mathbf{\sigma } \\ 
\mathbf{\sigma } & 0
\end{array}
) $, with $\mathbf{\sigma}$ being the Pauli matrix and $c$ the speed of light. 
It will also be instructive to introduce the sum (for each ${\bf k}$) of Berry curvatures over the occupied bands:
\begin{equation}
\Omega ^z ( \mathbf{k}) =\sum_nf_n\Omega ^z _n( 
\mathbf{k}).   \label{6}
\end{equation}
Then the intrinsic AHC is an integration over the Brillouin zone (BZ): 
\begin{equation}
\sigma  _{xy}=-\frac{e^2}\hbar \int_{BZ}\frac{%
d^3{k}}{( 2\pi ) ^3}\Omega ^z(\mathbf{k}). 
\label{7}
\end{equation}

The recent development of highly accurate \textit{ab initio} electronic structure 
calculation methods enables us to complete the work of Karplus and Luttinger by evaluating
their intrinsic Hall conductivity and comparing it with experiment. 
We employ the full-potential linearized augmented plane-wave 
method \cite{singh1994} with the generalized
gradient approximation (GGA) for the exchange-correlation potential \cite{perdew1996}.
Fully relativistic band calculations were performed
using the program package WIEN2K \cite{blaha2001}. A converged ground state with magnetization in the \mbox{$[$}001\mbox{$]$} direction 
was obtained 
using 20,000 {\bf k}
points in the first Brillouin zone and $K_{\max }R_{MT}=10$, 
where $R_{MT}$ represents the muffin-tin radius and $K_{\max }$ the maximum size of the reciprocal-lattice vectors. 
In this calculation, wavefunctions and potentials inside the
atomic sphere are expanded in spherical harmonics $Y_{lm}(\theta ,\phi )$ up to $l=10$ and $4$, respectively, and  
3s and 3p semi-core local orbitals are included in the basis set. 
The calculations were performed using the experimental lattice constant of $2.87$ $\mathrm{\AA }$. 
The spin magneton number was found to be $2.226$,
compared to the experimental value of $2.12$ as deduced from
measurements of the magnetization \cite{danan1968} and of the $g$ $(=2.09)$ factors.  The calculated energy bands are
shown in Fig.1, and are very similar to those obtained in Ref. \cite{singh1975}.  
If the spin-orbit interaction is parameterized as $\xi \, \mathbf{l\cdot s}$, its strength $\xi$ is found to be approximately $5.1$ mRy
from the band splitting near the H point and the Fermi energy.

\begin{figure}[!htb]
\begin{center}
\resizebox *{8.0cm}{6.18cm}{\includegraphics*{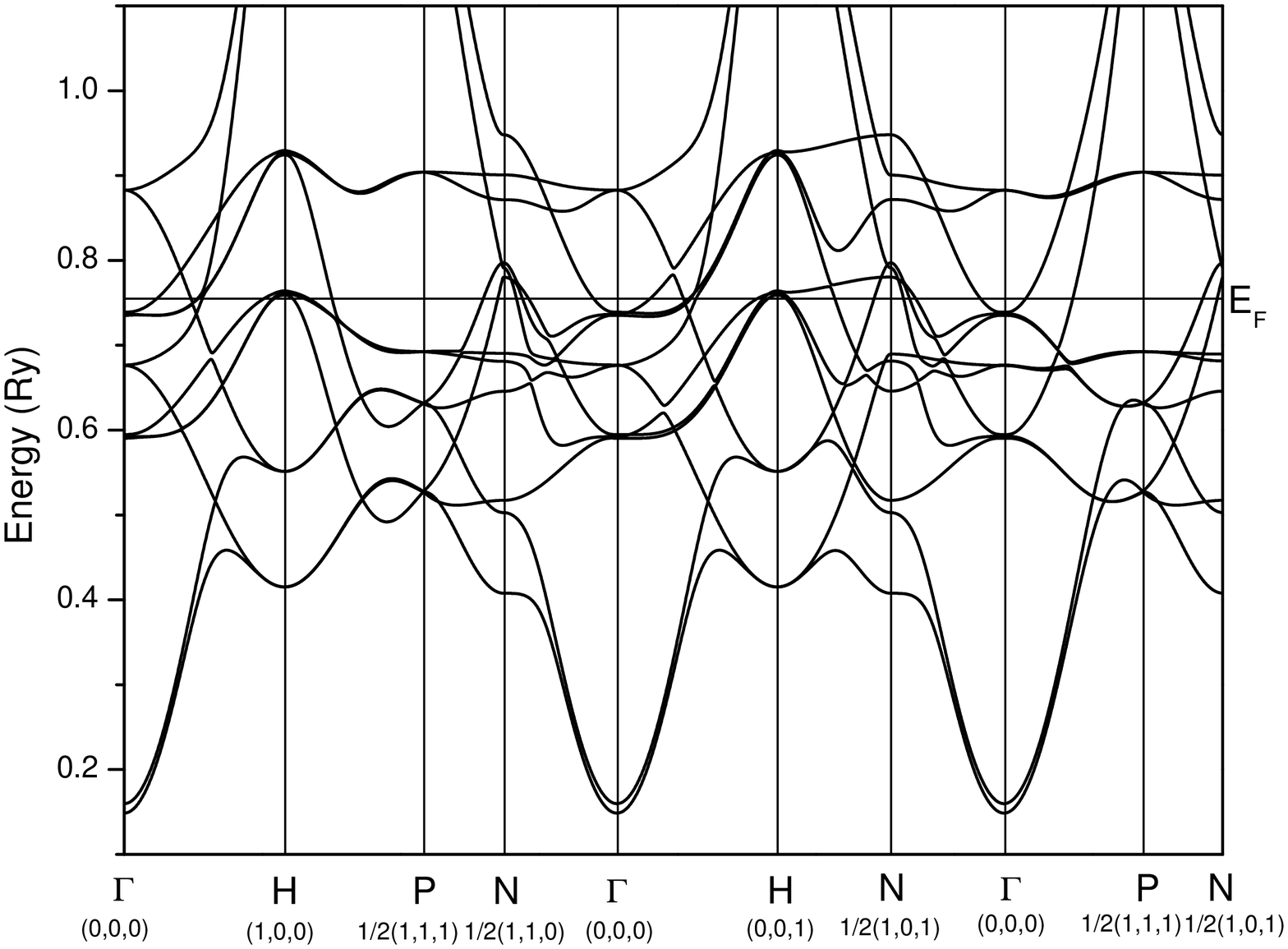}}
\end{center}
\caption{ Band structure of iron along high symmetry lines in the Brillouin
zone. The magnetization direction is along [001]. }
\label{fig:fig1}
\end{figure}

\begin{figure}[!htb]
\begin{center}
\resizebox *{8.0cm}{6.18cm}{\includegraphics*{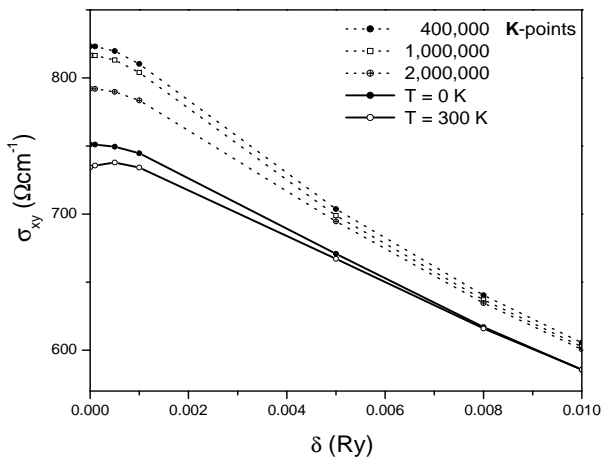}}
\end{center}
\caption{ Anomalous Hall effect vs. $\delta $ with different numbers of {\bf k} points in full Brillouin zone.
Here $\delta $ is introduced by adding $\delta ^2$ to the denominator in Eq.(\ref{5}).
The solid lines were obtained by an adaptive mesh refinement method. 
}
\label{fig:fig2}
\end{figure}

After obtaining the self-consistent potential with 20,000 \textbf{k} points, 
we calculated the Berry curvature with several larger sets of \textbf{k}-points
in order to achieve the convergence for $\sigma _{xy}$ shown in Fig. 2. 
The Monkhorst-Park special-point method \cite{monkhorst1976} was used for the integration in Eq.(\ref{7}).
To go beyond 2,000,000 points, we adopted a method of adaptive mesh refinement, 
{\it i.e.}, when $\Omega ^z({\mathbf k})$ is large at a certain {\bf k} point, we construct a finer
mesh by adding 26 additional points around it.  
This procedure yields a converged value of $\sigma _{xy}=751$ $\mathrm{(\Omega cm)^{-1}}$ at zero temperature
(using a step function for the 
Fermi-Dirac distribution) and a slightly smaller value of $\sigma _{xy}=734$ $\mathrm{(\Omega cm)^{-1}}$ at room temperature (300 K). 
Our result is in fair agreement with the value 1032 $\mathrm{(\Omega cm)^{-1}}$
extracted from Dheer's data on iron whiskers \cite{dheer1967} at room temperature.   

The slow convergence is caused by the appearance of large contributions to $\Omega ^z$ 
of opposite sign which occur in very small regions of 
\textbf{k}-space. Spin-orbit effects are small except when they mix states that would otherwise be degenerate 
or nearly degenerate, and even then, those
mixed states will contribute nearly canceling contributions to $\Omega ^z$. Only when the Fermi surface lies in a spin-orbit 
induced gap is there a large contribution.  This can be seen in Fig. 3 where the Berry curvature along lines in \textbf{k}-space
is compared with energy bands near $E_F$ and in Fig. 4 where it is compared with the intersection of the Fermi surface with
the central (010) plane in the Brillouin zone.

\begin{figure}[!htb]
\begin{center}
\resizebox *{8.0cm}{6.18cm}{\includegraphics*{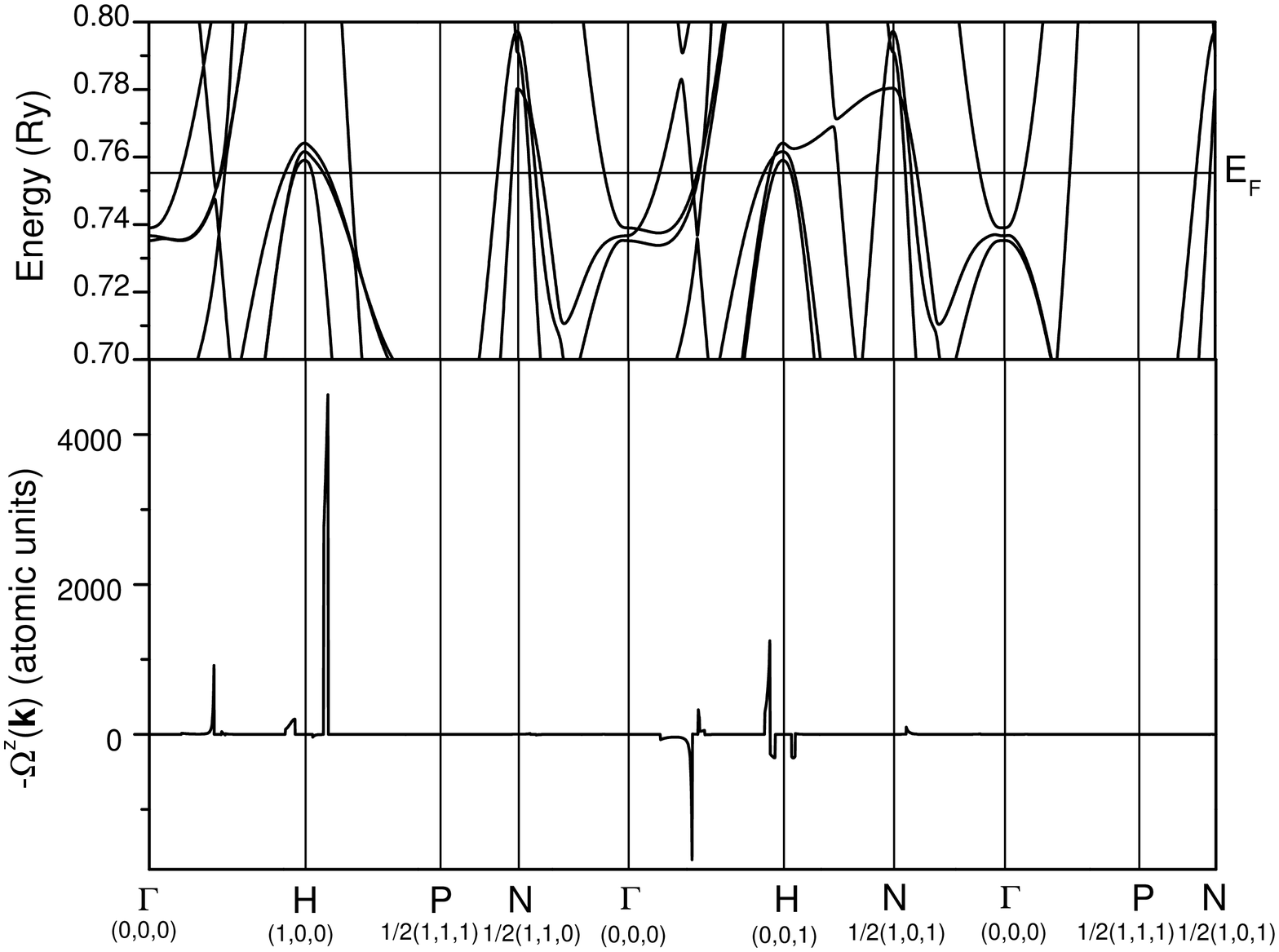}}
\end{center}
\caption{ Band structure near Fermi surface (upper figure) and Berry
curvature $\Omega ^z(\mathbf{k}) $ (lower figure) along symmetry lines. }
\label{fig:fig3}
\end{figure}

\begin{figure}[!htb]
\begin{center}
\resizebox *{7.8cm}{7.45cm}{\includegraphics*{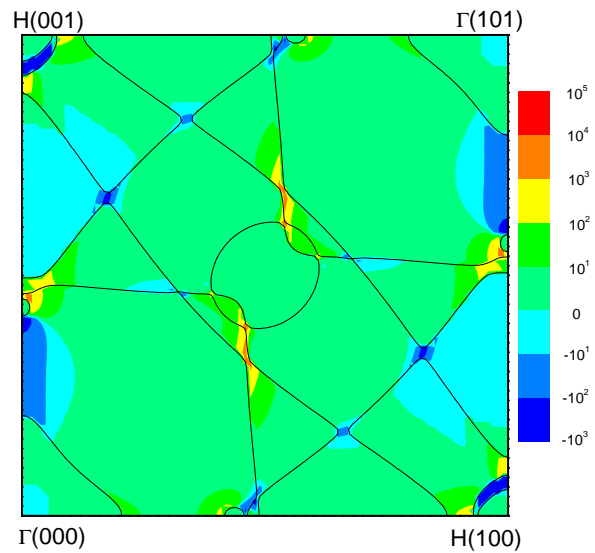}}
\end{center}
\caption{ (010) plane Fermi-surface (solid lines) and Berry curvature $-\Omega ^z
(\mathbf{k})$ (color map).  $-\Omega_z$ is in atomic units.}
\label{fig:fig4}
\end{figure}

In order to further understand the role of spin-orbit coupling in the AHE, we artificially varied
the speed of light, thereby changing the spin-orbit coupling strength $\xi \propto c^{-2}$.  
As shown in Fig. 5, $\sigma_{xy}$ is linear in $\xi$ for small coupling, 
but not for large coupling.  For iron, nonlinearities become significant for 
$\xi / \xi _0>1/2$, which means that the spin-orbit interaction in iron cannot be 
treated perturbatively.  

\begin{figure}[!htb]
\begin{center}
\resizebox *{8.0cm}{6.18cm}{\includegraphics*{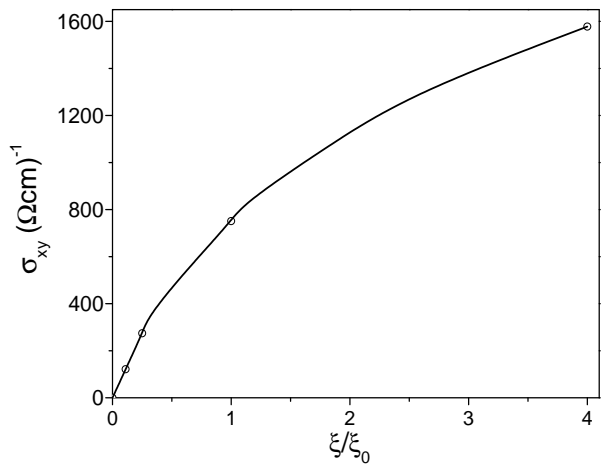}}
\end{center}
\caption{ Calculated anomalous Hall conductivity (open circles) vs. spin-orbit coupling $\xi $, $\xi _0
$ is spin-orbit coupling strength of iron. The line is a guide to the eye.}
\label{fig:fig5}
\end{figure}

So far we have discussed only the dc-AHE. It is straightforward to 
extend our calculation to the ac Hall case by using the Kubo formula \cite{sinova2003} approach:
\begin{eqnarray}
\sigma (\omega ) _{xy} &=& \frac{e^2}\hbar
\int_{V_G}\frac{d^3{k}}{( 2\pi ) ^3}\sum_{n\neq n^{\prime
}}( f_{n,\mathbf{k}}-f_{n^{\prime },\mathbf{k}})   \label{8} \\
&&\times \frac{\mathop{\rm Im}\left\langle \psi _{n\mathbf{k}}\left| v
_x\right| \psi _{n^{\prime }\mathbf{k}}\right\rangle \left\langle \psi
_{n^{\prime }\mathbf{k}}\left| v _y\right| \psi _{n\mathbf{k}%
}\right\rangle }{( \omega _{n^{\prime }}-\omega _n) ^2-
(\omega+i\epsilon)^2},  \nonumber
\end{eqnarray}
where $\epsilon$ is a positive infinitesimal.
In the upper panel in Fig.6, we show results for the imaginary part of $\omega \sigma_{xy}$ as a function of
frequency that are in agreement with earlier calculations\cite{mainkar1996}.  
Experimental results \cite{krinckik1967} are in excellent
agreement below $1.7$ $\mathrm{eV}$ but become smaller at higher energies.
In the lower panel of the figure, the real part of the Hall conductivity, obtained from the imaginary part by a 
Kramers-Kronig relation, is shown as a function of frequency.  The dc limit result,
$\sigma (\omega =0)_{xy}= 750.8$ $(\mathrm{\Omega cm)^{-1}}$, is essentially identical to that 
obtained from Eq.(6).  Despite the small discrepancy with theory in the dc limit, 
the experimental point $\bullet $ \cite{dheer1967} seems 
to agree rather well with the overall trend of the frequency 
dependence of the calculated AHC.

\begin{figure}[!htb]
\begin{center}
\resizebox *{8.0cm}{6.18cm}{\includegraphics*{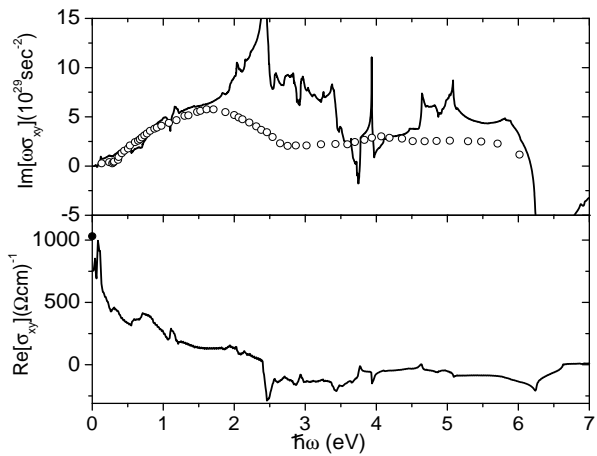}}
\end{center}
\caption{ Frequency dependence of the Hall conductivity at zero temperature.  In the upper panel, 
the calculated imaginary part of $\omega  \sigma  _{xy}$ (solid curve) is 
compared with experimental results $\circ $
Ref. (\protect\cite{krinckik1967}).  In the lower panel, the real part of $\sigma  _{xy}$ 
is shown together with the dc experiment value $\bullet $ extracted from Ref. (\protect\cite{dheer1967}). }
\label{fig:fig6}
\end{figure}

In conclusion we have shown that the AHC of bcc Fe, and presumably all other transition metal
ferromagnets, is primarily intrinsic.
(The only previous evaluation of the AHC of which we are aware \cite{leribaux1966} found that $\sigma _{xy}
=20.9 (\mathrm{\Omega cm})^{-1}$) for Fe.  The remaining discrepancy between theory and experiment is likely due to 
shortcomings of the GGA, neglect of scattering effects, and experimental uncertainties. 


This work was supported by DoE/DE-FG03-02ER45958, and by the Welch
Foundation. Y. Yao also acknowledges financial support from the NSF of
China (10134030, 60021403), the National Key Project for Basic Research
(G2000067103), and the National Key Project for High-Tech (2002AA311150).
L.K. was supported by NSF grant DMR-0073546.  AHM and JS acknowledge helpful
discussions with P. Bruno.

\end{document}